# Synchronization in counter-rotating oscillators

Sourav K. Bhowmick[1], Dibakar Ghosh[2], Syamal K. Dana[1]

[1]*Central Instrumentation, Indian Institute of Chemical Biology*
(*Council of Scientific and Industrial Research*)
*Jadavpur, Kolkata 700032, India*
[2]*Department of Mathematics, Dinabandhu Andrews College, Kolkata 700084, India.*

An oscillatory system can have opposite senses of rotation, clockwise or anticlockwise. We present a general mathematical description how to obtain counter-rotating oscillators from the definition of a dynamical system. A type of mixed synchronization emerges in counter-rotating oscillators under diffusive scalar coupling when complete synchronization and antisynchronization coexist in different state variables. We present numerical examples of limit cycle van der Pol oscillator and, chaotic Rössler and Lorenz systems. Stability conditions of mixed synchronization are analytically obtained for both Rössler and Lorenz systems. Experimental evidences of counter-rotating limit cycle and chaotic oscillators and, mixed synchronization are given in electronic circuits.

**Counter-rotating vortices coexist in a fluid medium, atmosphere and ocean, which destroy each other when they collide and reborn if the interaction is withdrawn. Attempts are found in literature to explain such spatio-temporal behaviors and their instabilities. On the other hand, counter-rotating vortices are created in physical systems, plasma flow and Bose-Einstein condensates, for useful purposes. The trajectory of nonlinear dynamical systems too shows opposite senses of rotation, clockwise and anticlockwise direction. A question naturally arises how counter-rotating vortices or oscillators are created and interact with each other. These issues are not well understood. We make an attempt to address the questions here, however, restricting our interest to the simpler cases of nonlinear dynamical systems. In studies of nonlinear oscillators, all oscillators are assumed to rotate in same direction, either clockwise or anticlockwise. These co-rotating oscillators when interact through coupling emerge into a synchronization regime where all the state variables follow one and unique correlation rule, complete synchronization or antisynchronization. In contrast, counter-rotating oscillators under diffusive coupling show emergence of an uncommon type of synchronization defined as mixed synchronization. In this synchronization regime, some of the state variables emerge into complete synchronization while others are in antisynchronization. We try to understand here the underlying principle how counter-rotating oscillators are created from a known dynamical system and then derive the stability condition of mixed synchronization. We construct counter-rotating oscillators in electronic circuits to evidence mixed synchronization.**

## I. Introduction

Synchronization is well investigated [1] in oscillatory systems, limit cycle as well as chaotic. The nature and strength of coupling decides the type of synchronization, either in both amplitude and phase [1-2] or only in phase [1, 3]. The emergent phase correlation in the coupled oscillators may be in-phase or antiphase. However, the sense of rotation of the trajectories of the interacting oscillators in these studies is always assumed same. This class of co-rotating oscillators emerge into a kind of coherent state, as example, complete synchronization (CS) or antisynchronization (AS). All the state variables follow same type of correlation, CS or AS. On the other hand, a recent study [4] revealed that the sense of rotation of the trajectory of a dynamical system may be clockwise or anticlockwise. Apparently, this sounds trivial and the topic remained ignored so far. But it is very nontrivial [4] that such counter-rotating oscillators under diffusive coupling emerge into a different type of correlation. The state variables of the coupled counter-rotating oscillators emerge into a mixed synchronization (MS) where a pair of state variables develops a CS state while another pair is in a AS state. It is true that existence of a MS state was reported earlier [5] in coupled co-rotating Lorenz systems for a specific scalar coupling. This emergent MS state has justification in the inherent axial symmetry of the Lorenz system but it was not observed, to our best knowledge, in other systems under linear diffusive coupling. However, a design of coupling was proposed [6] for engineering a MS state in co-rotating oscillators. In contrast, in counter-rotating oscillators, the MS state emerges naturally under diffusive scalar coupling. This observation encourages further investigation to understand the origin of counter-rotating oscillators and related coupled dynamics.

In fact, counter-rotating vortices (CRV) are the key elements of dynamical atmosphere and ocean [7]. They coexist in a fluid medium which destroy each other when they collide and reborn if the interaction is withdrawn. How a CRV pair is created? How they interact with each other? These are important questions necessitate appropriate answer for reducing damage of an aircraft, since CRV pair originates in the wake of an aircraft during take-off and landing [8]. On the other hand, counter-rotating vortices are deliberately created in physical systems, magnetohydrodynamics of plasma flow [9], Bose-Einstein condensates [10] for useful purposes. Counter-rotating spirals are also seen in biological medium such as protoplasm of the *Physarum* plasmodium [11]. All these facts strongly suggest that counter-rotating time evolving dynamical systems may exist in nature and may have beneficial role to play. From these viewpoints, it demands a special attention and a possible revision of the existing knowledge of synchronization, particularly, in context of collective behaviors of a network of a mixed population of counter-rotating oscillators. Presently, we restrict our study to a simpler case of two counter-rotating oscillators.

The existence of clockwise and counter-clockwise rotation in a dynamical system was noticed by Tabor [12] in relation to rotation of a 2D limit cycle system near the Hopf bifurcation point. However, the variation of rotational direction was not considered while investigating collective dynamics of oscillators until recently [4]. The study made an extension to chaotic oscillators. Any general statement was still missing how to induce counter-rotation and to explain the mechanism of counter-rotation in a known dynamical system. We present a general mathematical description, in this paper, how to create counter-rotating motion in a given dynamical system, limit cycle or chaotic, and then investigate the collective behavior. A particular type of MS regime emerges in pairs of state variables for different choices of scalar coupling. We analytically established the stability conditions of MS for coupled chaotic systems. We tested our proposition using numerical simulations of limit cycle as well as chaotic system. Finally, we supported the numerical results with experiments by constructing electronic circuits of counter-rotating van der Pol and piecewise Rössler oscillators.

The rest of the paper is organized as follows: theory of counter-rotating oscillators is presented in section II with examples in section IIA. Mixed synchronization in van der Pol oscillator [13], chaotic Rössler oscillator [14] and Lorenz systems [15] is described in section III. The stability condition of MS in the chaotic systems are described in section IV. Experimental evidences of counter-rotating oscillators and MS regime is described in V with a conclusion in section VI.

## II. Counter rotating oscillators: Method

A dynamical system can always be expressed by,

$$\dot{X} = AX + f(X) + C \quad (1)$$

where $X \in R^n$ is a state vector, $A$ is a $n \times n$ constant matrix and represents the linear part of the system, $f: R^n \rightarrow R^n$ contains the nonlinear part of the system and, $C$ is a $n \times 1$ constant matrix. For a 3D system when $A=(a_{ij})_{3\times 3}$ and $X=[x, y, z]^T$, a general procedure is stated in the following steps for deriving counter-rotating oscillators. Crucial changes are made in the $A$ matrix, i.e., in the linear part of the system,

i) there must exist at least one pair of non-zero conjugate elements $a_{ij}, a_{ji}, i \neq j$ in the matrix $A$, where $i, j$=1,2,3.

ii) then replace the element $a_{ij}$ by $-a_{ij} (i \neq j)$ or vice versa.

The clockwise or anticlockwise rotation of the trajectory of a 3D dynamical system is described as the direction of rotation of the trajectory in a 2D plane in respect to the third coordinate. As stated in (ii), the sense of rotation of a trajectory can be flipped [4] by altering the sign of the conjugate pair of elements of the $A$ matrix. This is as usually done in case of quantum rotation [10] by changing the sign of the angular frequency. This can be extended to frame a general rule following the standard Eurler's rotation theorem [16]. In a 3D system, as stated in the Euler's rotation theorem, any rotation can be described using three angles of rotation around three coordinate axes. The final rotation matrix is a product of three rotational matrices and given by,

$$A = \begin{bmatrix} a_{11} & a_{12} & a_{13} \\ a_{21} & a_{22} & a_{23} \\ a_{31} & a_{32} & a_{33} \end{bmatrix} \quad (2)$$

where
$a_{11} = \cos\varphi\cos\kappa$ ; $a_{12} = \cos\omega\sin\kappa + \sin\omega\sin\varphi\cos\kappa$
$a_{13} = \sin\omega\sin\kappa - \cos\omega\sin\varphi\cos\kappa$
$a_{21} = -\cos\varphi\sin\kappa$ ; $a_{22} = \cos\omega\cos\kappa - \sin\omega\sin\varphi\sin\kappa$
$a_{23} = \sin\omega\cos\kappa + \cos\omega\sin\varphi\sin\kappa$
$a_{31} = \sin\varphi$ ; $a_{32} = -\sin\omega\cos\varphi$ ; $a_{33} = \cos\omega\cos\varphi$

and $\kappa, \varphi, \omega$ are the rotation angles in any of the 2D coordinate spaces around the remaining third coordinate. Suppose we consider the $x$-$y$ plane of rotation around the $z$-axis then the rotation angle is $\kappa$ ($\omega = 0$, $\varphi = 0$): the new co-ordinate ($x_1, y_1, z_1$) of the point ($x, y, z$) after rotation by the angle $\kappa$ is written,

$$\begin{bmatrix} x_1 \\ y_1 \\ z_1 \end{bmatrix} = \begin{bmatrix} \cos\kappa & \sin\kappa & 0 \\ -\sin\kappa & \cos\kappa & 0 \\ 0 & 0 & 1 \end{bmatrix} \begin{bmatrix} x \\ y \\ z \end{bmatrix} \quad (3)$$

The direction of rotation will reverse if the rotation angle $\kappa$ is taken as negative i.e., if $\kappa$ is substituted by -$\kappa$. The transformed coordinate ($x_2, y_2, z_2$) in reverse direction is,

$$\begin{bmatrix} x_2 \\ y_2 \\ z_2 \end{bmatrix} = \begin{bmatrix} \cos\kappa & -\sin\kappa & 0 \\ \sin\kappa & \cos\kappa & 0 \\ 0 & 0 & 1 \end{bmatrix} \begin{bmatrix} x \\ y \\ z \end{bmatrix} \quad (4)$$

A change of sign of the elements $a_{12}, a_{21}$ in (4) is thus needed to induce a change in the direction of rotation in the $x$-$y$ plane in respect of the $z$-direction. The counter-rotation in the $y$-$z$ plane in respect to the $x$-axis is obtained, $\omega \neq 0$ ($\kappa = 0$, $\varphi = 0$),

$$\begin{bmatrix} x_1 \\ y_1 \\ z_1 \end{bmatrix} = \begin{bmatrix} 1 & 0 & 0 \\ 0 & \cos\omega & \sin\omega \\ 0 & -\sin\omega & \cos\omega \end{bmatrix} \begin{bmatrix} x \\ y \\ z \end{bmatrix} \quad (5)$$

The sign of the elements $a_{23}$ and $a_{32}$ of the $A$ matrix are to be exchanged for counter-rotation in the $y$-$z$ plane with respect to the $x$-axis. Similarly, we can derive a rotation matrix for rotation angle $\varphi \neq 0$ ($\kappa = 0$, $\omega = 0$) in the x-z plane around the y-axis and accordingly introduce an exchange sign in $a_{13}$ and $a_{31}$ elements to obtain counter-rotation. The general matrix (2) is thus obtained by multiplication of the three rotational matrices for three angles of rotation, $\omega, \varphi, \kappa$. We consider only one axis of counter-rotation for our application and hence suffice to use (3) for our studies. We elaborate this with real examples of dynamical systems.

**IIA. Counter rotating oscillators: Examples**

Here we describe examples of counter-rotating oscillators,

(1) Chua oscillator [17]
$$\dot{x}_i = a(x_i - \omega_i y_i - g(x_i)), \ \dot{y}_i = b(\omega_i x_i - y_i + z_i)$$
$$\dot{z}_i = -cy_i \quad (6)$$

where
$$g(x_i) = m_1 x_i + m_1 - m_0 \quad \text{if } x_i \leq -1$$
$$= m_0 x_i \quad \text{if } -1 \leq x_i \leq 1 \text{ and}$$
$$= m_1 x_i - m_0 - m_1 \quad \text{if } 1 \leq x_i$$

$a = 15.6$, $b = 1$, $c = 33$, $m_0 = -8/7$, $m_1 = -5/7$, $i=1,2$, and $\omega_1=1$, $\omega_2=-1$.

(2) Sprott system [18]
$$\dot{x}_i = x_i y_i - \omega_i z_i, \ \dot{y}_i = x_i - y_i$$
$$\dot{z}_i = \omega_i x_i + 0.3 z_i \quad (7)$$

where $i=1, 2$ and $\omega_1=1$, $\omega_2=-1$

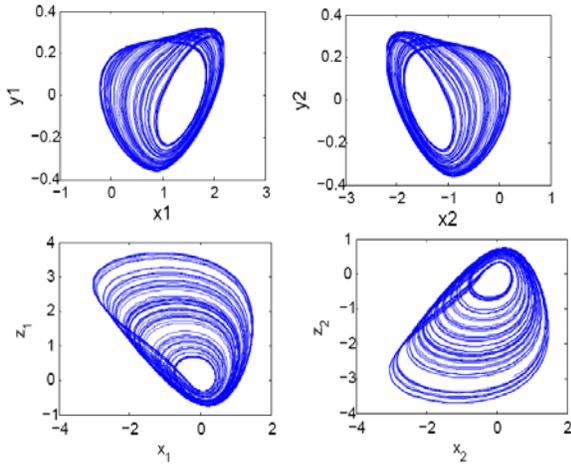

Fig.1. Phase portraits of counter-rotating oscillators. Chua system (6) in upper two panels, and Sprott system (7) in lower two panels.

Phase portraits of counter-rotating Chua oscillator and a Sprott system are shown in Fig.1. Clearly by changing the sign of a conjugate pair of elements in the linear matrices of the systems, a counter-rotation in the trajectories of the dynamical systems are induced. Note that the rotational plane of the Chua model in the upper panels is the x-y plane for opposite signs of $a_{12}$ and $a_{21}$ elements in the linear matrix of (6). While in Sprott system in the lower panels, it is the x-z plane when the $a_{13}$ and $a_{31}$ elements of the linear matrix of (7) are altered. Noteworthy that two largest Lyapunov exponents of two oscillators before coupling are almost same [4] for $\omega_1 = 1$ and $\omega_2 = -1$. Note that, in addition to changes in the sense of rotation, there is a change in the position of the attractors in phase space (Fig.1).

**III. Synchronization of counter-rotating oscillators**

We consider two coupled counter-rotating oscillators to explore synchronization,

$$\dot{x} = Ax + f(x) + C + HG(y, x) \quad (8)$$
$$\dot{y} = A'y + f(y) + C + HG(x, y) \quad (9)$$

where $x=[x_1 \ y_1 \ z_1]^T$, $y=[x_2 \ y_2 \ z_2]^T$ are the state vectors, $A$ and $A'$ are the linear matrices for counter-rotating oscillators and, $H=diag(\varepsilon_1, \varepsilon_2, \varepsilon_3)$ is the coupling matrix,

$$\text{and} \quad G(x, y) = \begin{bmatrix} x_1 - x_2 \\ y_1 - y_2 \\ z_1 - z_2 \end{bmatrix}.$$

When two counter-rotating oscillators are coupled in a scalar mode, a MS state emerges above a critical coupling. The scalar coupling must use at least one of the pairs of variables from the plane of rotation (x-y plane), either the pair of $x_1$-$x_2$ variables ($\varepsilon_1=\varepsilon$, $\varepsilon_2=0$, $\varepsilon_3=0$) or the pair of $y_1$-$y_2$ variables ($\varepsilon_1=0$ $\varepsilon_2=\varepsilon$, $\varepsilon_3=0$) of the coupled oscillators. Alternatively, if one oscillator is rotating clockwise in the $x_1$-$z_1$ plane and another in anticlockwise direction in the $x_2$-$z_2$ plane, either of the couplings, ($\varepsilon_1=\varepsilon$, $\varepsilon_2=0$, $\varepsilon_3=0$) and ($\varepsilon_1=0$, $\varepsilon_2=0$, $\varepsilon_3=\varepsilon$), can be used for MS and so on. We elaborate first using a limit cycle van der Pol oscillator,

$$\dot{x}_1 = y_1, \ \dot{y}_1 = b(1-x_1^2)y_1 - x_1 \quad (10)$$

where $b$ is the only parameter. The linear matrix of (10) for clockwise rotation is given by $A$ and a counter-clockwise rotation is derived by replacing it with $A'$,

$$A = \begin{bmatrix} 0 & 1 \\ -1 & b \end{bmatrix}; \quad A' = \begin{bmatrix} 0 & -1 \\ 1 & b \end{bmatrix} \quad (11)$$

Two counter-rotating van der Pol oscillators under a scalar diffusive coupling are,

$$\dot{x}_i = \omega_i y_i + \varepsilon(x_j - x_i); \quad \dot{y}_i = b(1-x_i^2)y_i - \omega_i x_i \quad (12)$$

where $i, j=1,2$ and $\omega_1=1$, $\omega_2=-1$, $\varepsilon$ is the coupling strength.

Numerically simulated phase portraits show two isolated counter-rotating oscillators in Fig.2(a) and 2(b) where the rotation of the trajectory is made opposite in the x-y plane by changing the sign of $\omega_1$. As stated above, a MS state only emerges in two different state variables as shown in Fig.2(c) and 2(d). The $x_1$ vs. $x_2$ plot confirms a CS state when $x_1$ and $x_2$ are only directly coupled while $y_1$ vs. $y_2$ shows an AS state. The coupling has a critical value for a choice of system parameter $b$ as shown in a phase diagram in Fig.2(e). The boundary of the MS state in a shaded region and nonsynchronization in the white region is the critical coupling.

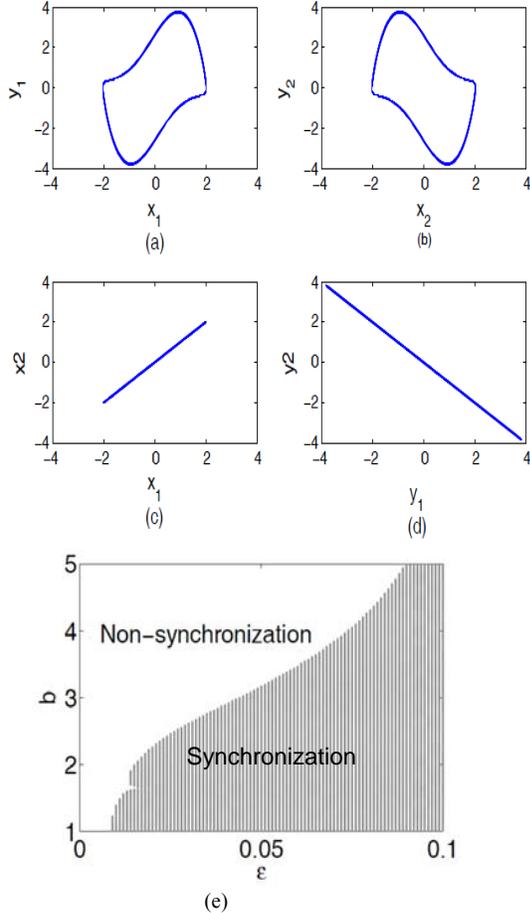

Fig 2: Counter-rotating van der Pol oscillators, $b=2.0$. Phase portraits of uncoupled van der Pol oscillators, (a) linear matrix $A$, (b) linear matrix $A'$. MS for $\varepsilon=0.075$, (c) CS, $x_1$ vs. $x_2$, (d) AS, $y_1$ vs. $y_2$. Phase diagram in $\varepsilon$-$b$ plane showing MS boundary in (e).

Next we use the Rössler oscillator and the Lorenz oscillator to elaborate counter-rotating chaotic systems and their MS behavior. As a first example, we consider the Rössler oscillator,

$$\dot{x} = -\omega_1 y - z, \quad \dot{y} = \omega_1 x + ay, \quad (13)$$
$$\dot{z} = b + z(x-c)$$

which is chaotic for $a=b=0.2$, $c=10$, $\omega_1=1$. The linear matrices of the Rössler model for counter-rotations,

$$A = \begin{bmatrix} 0 & -\omega_1 & -1 \\ \omega_1 & a & 0 \\ 0 & 0 & c \end{bmatrix}; \quad A' = \begin{bmatrix} 0 & \omega_1 & -1 \\ -\omega_1 & a & 0 \\ 0 & 0 & c \end{bmatrix} \quad (14)$$

where the sign of the $a_{12}$, $a_{21}$ elements in the linear matrix $A$ are altered.

The counter-rotating Rössler oscillators after coupling via a single variable or scalar coupling is

$$\dot{x}_i = -\omega_i y_i - z_i + \varepsilon(x_j - x_i)$$
$$\dot{y}_i = \omega_i x_i + ay_i \quad (15)$$
$$\dot{z}_i = b + z_i(x_i - c)$$

where $i=1, 2$ represents two oscillators, $\omega_1 = 1$, $\omega_2 = -1$ and $\varepsilon$ is the coupling strength as usual. Phase portraits of counter-rotating Rössler oscillators before coupling are shown in Figs.3(a) and 3(b). For scalar diffusive coupling, a MS state emerges for $\varepsilon \geq \varepsilon_c$, a critical coupling. To realize MS, the oscillators must be coupled at least by one of the state variables involved in the plane of rotation. As described above, CS is noticed in that pair of variables which are directly coupled, and AS in the other pair of variables those are not directly coupled in the rotational plane. The coupling of the counter-rotating Rössler oscillators is made via $x$-variables, hence $x$-variables are in a CS state and $y$-variables (since the plane of rotation is $x$-$y$) are in an AS state. The third variables $z_1$-$z_2$ are in a CS state (not shown here). This third pair of variables may be in CS or AS state, which is arbitrarily decided and not clearly understood so far. However it depends upon the system's inherent property too as found [5] in case of co-rotating Lorenz systems.

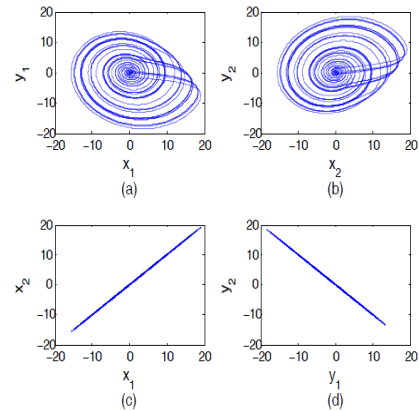

Fig.3. Counter-rotating Rössler oscillators: (a) linear matrix $A$, (b) linear matrix $A'$, (c) CS in $x_1$-$x_2$ and, (d) AS in $y_1$-$y_2$. Coupling strength, $\varepsilon = 0.15$.

Next, we consider the second example of two coupled counter-rotating Lorenz systems,

$$\dot{x}_i = \sigma(\omega_i y_i - x_i), \quad \dot{y}_i = \omega_i r\, x_i - y_i - x_i z_i + \varepsilon(y_j - y_i)$$
$$\dot{z}_i = x_i y_i - b z_i \quad (16)$$

where $\sigma$=10, $r$=28, $b$=8/3 and, $i, j = 1, 2$; $\omega_1$=1, $\omega_2$=-1. The linear matrices for counter-rotations of the isolated Lorenz system are $A$ and $A'$,

$$A = \begin{bmatrix} -\sigma & \sigma & 0 \\ r & -1 & 0 \\ 0 & 0 & -b \end{bmatrix} \text{ and } A' = \begin{bmatrix} -\sigma & -\sigma & 0 \\ -r & -1 & 0 \\ 0 & 0 & -b \end{bmatrix} \quad (17)$$

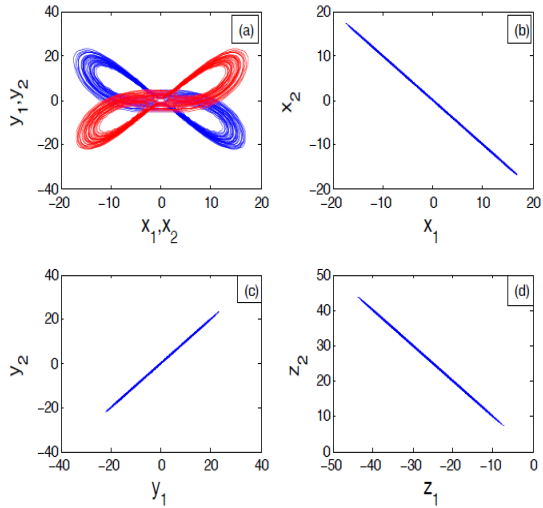

Fig.4. Counter-rotating Lorenz systems: attractors for (a) $A$ (online blue) and $A'$ (online red) in uncoupled state, (c) AS in $x_1$ vs.$x_2$, (c) CS in $y_1$ vs.$y_2$ and (d) AS in $z_1$ vs. $z_2$ and $\varepsilon = 1.5$.

The counter-rotating Lorenz attractors in uncoupled state are shown in Fig.4(a). When they are coupled as defined in (16), a MS scenario emerges as shown in Figs.4(b) and 4(c). They confirm CS in the $(y_1, y_2)$ pair, AS in the $(x_1, x_2)$ pair as expected for the coupling via $y$-$y$ variables. The $(z_1, z_2)$ pair is in an AS state whereas in case of Rössler oscillators we find them in a CS state. As mentioned above, MS was reported earlier [5] in two co-rotating Lorenz systems for $z$-$z$ scalar coupling above a critical value. There the $z$-$z$ variables were in an AS state when both the $x$-$x$ and $y$-$y$ variables were in CS. The reverse is also true [19]. In case of counter-rotating Lorenz oscillators, it is a different scenario.

**IV. Stability of Mixed Synchronization**

Now we explore the stability of MS in both counter-rotating Rössler and Lorenz systems, using linear stability analysis.

Let $\eta_1$ and $\eta_2$ represent the deviation from a synchronized state, their dynamics is governed by linear equations

$$\dot{\eta}_1 = A\eta_1 + f'(x)\eta_1 + H(\eta_2 - \eta_1) \quad (18)$$
$$\dot{\eta}_2 = A'\eta_2 + f'(y)\eta_2 + H(\eta_1 - \eta_2) \quad (19)$$

It is very difficult to analyze the stability of the system (18)-(19) and hence make an approximation [20] by taking time average of the *Jacobians*, $f'(x)$ and $f'(y)$ and replacing them by a constant $\lambda$. For MS in counter-rotating systems, $x=\pm y$ (CS or AS), let $\eta=\eta_1-\eta_2$,

$$\dot{\eta} = A\eta + \lambda\eta - 2H\eta = [A + \lambda I_3 - 2H]\eta \quad (20)$$

$\eta$=0 will be stable if $P = A + \lambda I_3 - 2H < 0$, where $I_3$ is a 3×3 identity matrix and the constant $\lambda$ can be chosen arbitrarily. The synchronized state is defined by $\eta = \eta_1 - \eta_2 = 0$, i.e., $x \pm y =$ constant. Numerical simulations show this constant as zero.

For two counter-rotating coupled Rössler oscillators,

$$P = \begin{bmatrix} \lambda - 2\varepsilon & -1 & -1 \\ 1 & a + \lambda & 0 \\ 0 & 0 & \lambda - c \end{bmatrix} \quad (21)$$

and assuming scalar coupling, $H = diag[\varepsilon\ 0\ 0]$, the eigenvalues of the $P$ matrix are,

$\lambda_1 = \lambda - c$,

$$\lambda_{2,3} = \frac{-(2\varepsilon - 2\lambda - a) \pm \sqrt{(2\varepsilon - 2\lambda - a)^2 - 4\{(\lambda + a)(\lambda - 2\varepsilon) + 1\}}}{2}$$

The stability criteria for a MS state are then defined by, all eigenvalues have negative real parts,

(i) $\lambda < c$,
(ii) if $(2\varepsilon - 2\lambda - a)^2 < 4\{(\lambda + a)(\lambda - 2\varepsilon) + 1\}$, $\lambda_{2,3}$ are complex and the stability condition is $2\varepsilon > 2\lambda + a$.
(iii) if $(2\varepsilon - 2\lambda - a)^2 > 4\{(\lambda + a)(\lambda - 2\varepsilon) + 1\}$, $\lambda_{2,3}$ are real and stability condition becomes $2\varepsilon > 2\lambda + a$ and $2\varepsilon < \lambda + 1/(\lambda + a)$.

The critical coupling for Rössler system shows two limits,

$$\lambda + a/2 < \varepsilon < \tfrac{1}{2}[\lambda + 1/(\lambda + a)] \quad (22)$$

For a choice of $\lambda = 0.025$ and system parameters, $a = b = 0.2$, $c = 10$, the critical coupling for MS stability in counter-rotating Rössler system is determined by (22) as $0.125 < \varepsilon < 2.2347$. The range of critical coupling varies with parameter $a$. In coupled counter-rotating Rössler oscillator, it is known *a priori*, from numerical simulations for coupling via $x$-variable, that the $x$-$x$ variables develop CS while the $y$-$y$ variables and the $z$-$z$ variables are in CS, the MS error is then defined by,

$e = \sqrt{\langle (x_2 - x_1)^2 + (y_2 + y_1)^2 + (z_2 - z_1)^2 \rangle}$ and it is calculated for a set of parameter $(a, \varepsilon)$ in ranges, $a \in [0.1, 0.3]$ and $\varepsilon \in [0, 4.5]$. The stable MS region is obtained numerically as indicated by dots (online blue dot) in Fig.5(a) with lower and upper boundaries. In Fig.5(a), the analytic lower threshold line is given by, $\varepsilon = \lambda + a/2$ and the upper threshold line is $\varepsilon = 1/2[\lambda + 1/(\lambda+a)]$. The analytic results closely match the numerical results of critical coupling. Similarly, for the Lorenz system, the eigenvalues of the $P$ matrix are obtained,

$$\lambda_1 = \lambda - b$$
$$\lambda_{2,3} = \frac{-(\sigma - 2\lambda + 1 + 2\varepsilon)}{2} \quad (23)$$
$$\pm \frac{\sqrt{(\sigma - 2\lambda + 1 + 2\varepsilon)^2 - 4\{(\sigma - \lambda)(1 + 2\varepsilon - \lambda) - r\sigma\}}}{2}$$

The stability criteria for the MS state becomes

$$\varepsilon > \max\{\lambda - \frac{1+\sigma}{2}, \frac{1}{2}(\lambda - 1 + \frac{r\sigma}{\sigma - \lambda})\}, \text{ where } \lambda \text{ satisfy } \lambda < b.$$

For a choice of $\lambda = -10.177$, the critical coupling curve is

$$\varepsilon_c = \frac{1}{2}(\lambda - 1 + \frac{r\sigma}{\sigma - \lambda}). \quad (24)$$

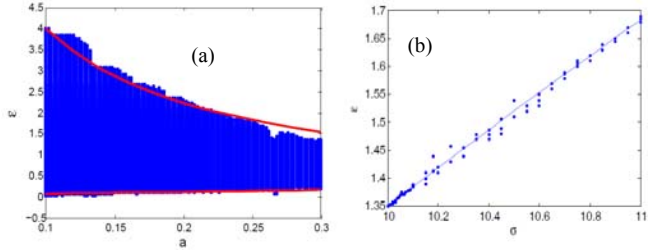

Fig.5.MS regime for (a) Rössler oscillator, (b) Lorenz system.

In Fig.5(b), the diagonal line is the analytic critical coupling defined by (24) and the solid circles are numerically simulated critical coupling which closely match. The analytical results on stability of MS in both the counter-rotating Rössler oscillators and Lorenz oscillator are further confirmed by estimating the transverse Lyapunov exponents. A necessary condition for synchronization to be stable is that the maximum transversal Lyapunov exponent $\lambda_{max}$ be negative [21]. The transversal Lyapunov exponent for the MS is determined using the variational equation of counter-rotating Rössler systems,

$$\dot{e}_1 = -2\varepsilon e_1 - e_2 - e_3, \quad \dot{e}_2 = e_1 + 0.2 e_2 \quad (25)$$
$$\dot{e}_3 = z_1 + e_1 + (x_2 - \mu)e_3$$

where $e_1 = x_1 - x_2$, $e_2 = y_1 + y_2$ and $e_3 = z_1 - z_2$ are the transversal deviation from the synchronization manifold.

In Fig.6(a), the $\lambda_{max}$ is plotted for coupled counter-rotating Rössler system with coupling strength $\varepsilon$. Two stability thresholds $\varepsilon_1$ and $\varepsilon_2$ are again found for MS. The MS state is stable in the interval $(\varepsilon_1, \varepsilon_2)$ where $\lambda_{max}<0$. The analytic result is approximately equal to the numerical result $0.125< \varepsilon <2.2732$. The critical coupling range for stability of MS in counter-rotating Rössler oscillators is similar [22] to its co-rotating counterpart. Stability criterion for coupled counter-rotating Lorenz systems is similarly obtained in Fig.6(b) where $\lambda_{max}$ is found negative above one critical coupling, $\varepsilon \gtrsim 1.35$ when MS emerges.

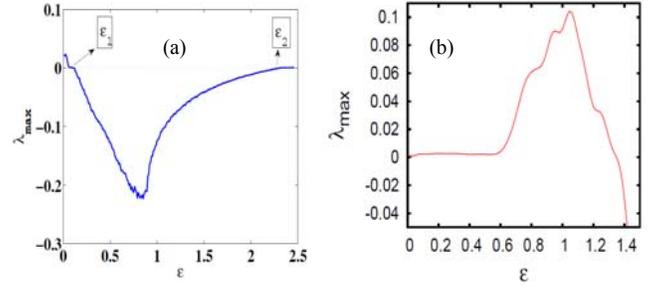

Fig.6.Maximum transversal Lyapunov exponent.(a) Rössler oscillator with two critical coupling, $\varepsilon_1=0.125$, $\varepsilon_2=2.2732$, (b) Lorenz oscillator with one critical coupling, $\varepsilon=1.135$.

**V. Mixed Synchronization: Experiment**

Finally we present experimental evidences of MS scenario in limit cycle as well as chaotic system. We first design an electronic analog of the limit cycle van der Pol model (12). We construct two counter-rotating van der Pol oscillators whose schematic circuits are shown in Fig.7. The first oscillator (oscillator-1) uses two integrators (U1-U2) to simulate output voltages representing the state variables $x_2$ and $y_2$, two multipliers (U4-U5) to simulate the cubic nonlinearity in (12). Both the integrators U1-U2 are in inverting mode. The U2 integrator also plays the role of an adder amplifier. The unity gain inverting amplifier U3 is used to make a necessary change in the sign of the voltage output of the inverting integrator U2 and then fed back to both U1 and U2 again. The gain of U3 defines the $\omega$–value which is considered unity here ($\omega=1$). A counter-rotating oscillator (oscillator-2) is similarly designed using two integrators (U6, U8), two multipliers (U10, U11) and an inverting amplifier U9. An additional inverting amplifier U7 is used to induce a reverse rotation (to define a value of $\omega=-1$) relative to the first one. Note that the output of U1 is directly connected to U2 in the oscillator-1 while, in the oscillator-2, U6 is connected to U8 via a unity gain inverting amplifier U7 to invert the sign of $\omega$. The coupling circuit is shown at the bottom of Fig.7. The coupling strength $\varepsilon$ is controlled by the resistances $R_{27}/R_{19}=R_{28}/R_{25}$. Output leads OUT-1 and OUT-2 are connected to the inverting inputs of U1 and U6 respectively to realize bidirectional coupling. Figure 8 shows the oscilloscope (Yokogawa DL 9140, 4-channel, 1GHz, 5GS/s) pictures of counter-rotating limit cycle attractors (upper row) as plots of $x_2$ vs. $y_2$ and $x_1$ vs. $y_1$ using output voltages from (U1, U2) and (U6, U8) respectively.

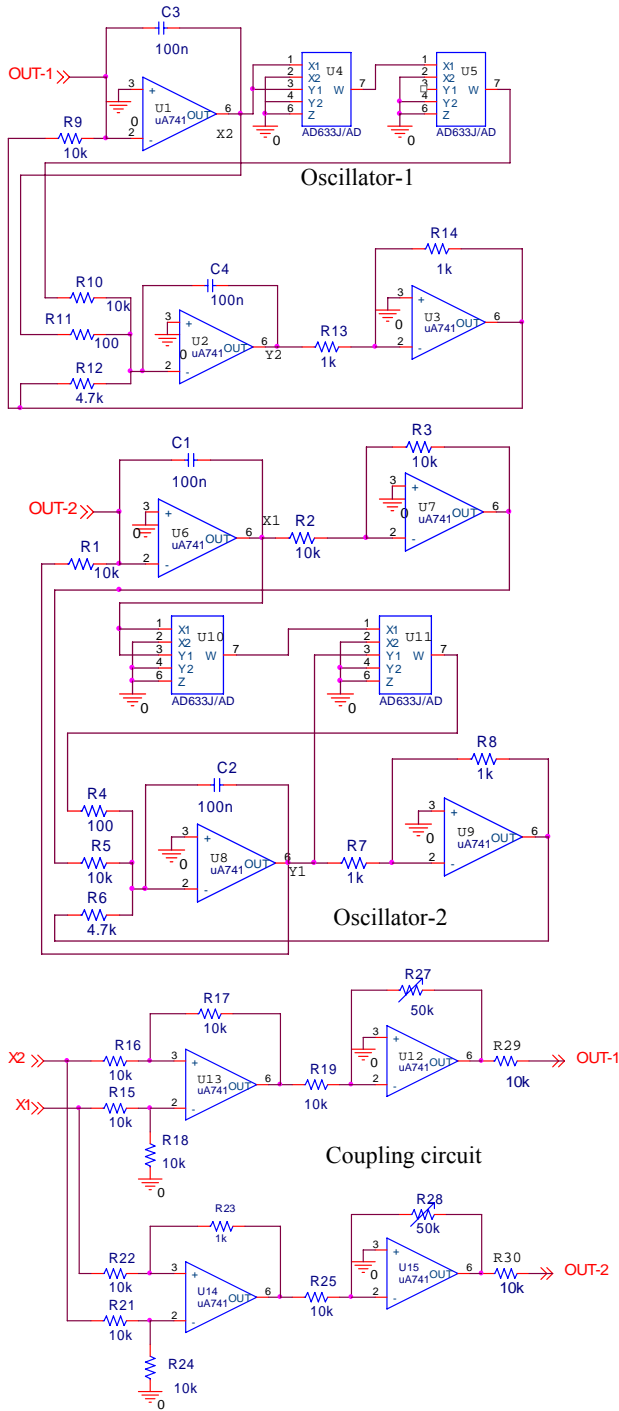

Fig.7. Counter-rotating van der Pol circuits. Oscillator-1 is the upper circuit ($\omega_1=+1$), oscillator-2 in the middle ($\omega_2=-1$). Coupling circuit is at the bottom. Oscillators show MS for $R_{27}=R_{28}$ larger than 20k$\Omega$.

All the measured time series are separately shown in next four lower rows. Time series as measured from U1 and U6 and shown in second row of $x_2$ (online red) and third row of $x_1$ (online yellow) respectively, are in CS and, as measured from U2 and U8 and shown in fourth row of $y_2$ (online green) and fifth row of $y_1$ (online blue) respectively, are in AS. In the bottom panels, $x_1$ vs. $x_2$ plot at left confirms CS scenario and, $y_1$ vs. $y_2$ plot at right reveals AS and thereby confirm the existence of a MS regime.

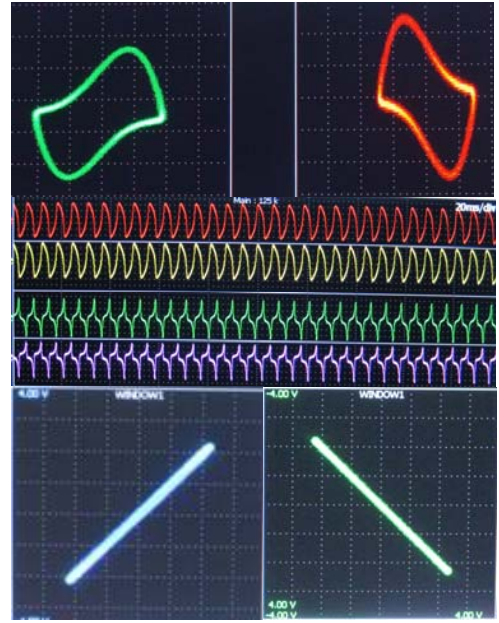

Fig.8. Oscilloscope pictures: counter-rotating attractors (online green and red) in upper row. Time series for coupled counter-rotating oscillators: $x_1$ in second row (online red) and $x_2$ in third row (online yellow) in CS, $y_1$ in fourth row (online green) and $y_2$ in fifth row (online blue) are in AS. Sixth row plots ($x_1$vs.$x_2$) show CS at left, AS ($y_1$vs.$y_2$) at right.

For experimental verification of MS scenario in a chaotic system, we use an electronic analog [23] of the piecewise Rössler model,

$$\dot{x} = -\alpha x - \omega\beta y - \lambda z$$
$$\dot{y} = \omega x + \gamma y - 0.02z \qquad (26)$$
$$\dot{z} = g(x) - z,$$
$$\text{where } g(x) = 0 \qquad \text{for } x \leq 3$$
$$= \mu(x-3) \qquad \text{for } x \geq 3$$

and $\alpha = 0.05$, $\beta = 0.5$, $\lambda = 1.0$, $\mu = 7$, $R_4$=10k$\Omega$, $\gamma = \dfrac{R_4}{R_{13}}$.

Next we construct two counter-rotating circuits of the piecewise linear Rössler model as shown in Fig.9. The oscillator-1 (upper circuit) is for $\omega$=1 and the oscillator-2 is for $\omega$=-1. All three integrators (U1, U3 and U5) in oscillator-1 are in inverting mode representing three dynamical equations in (26). The U2 is a simple inverting amplifier. The op-amp U4 with a diode D1 constructs the piecewise linear function. The counter-rotating oscillator (oscillator-2) is constructed in a similar manner using integrators (U6, U8 and U10) and inverting amplifier U7. The op-amp U9 and a diode is used to design the piecewise linear function for oscillator-2. The additional op-amp U11 is used to reverse the rotation of the oscillator-2. The gain of U11 is again considered unity since $\omega=-1$. For larger $\omega$, the gain can be set at a desired value. The $R_{13}$ is set at 49 kΩ to obtain chaotic dynamics. The coupling circuit is shown in Fig.10. The counter rotating attractors are observed using the oscilloscope (Yokogawa DL 9140) as shown in Fig.11.

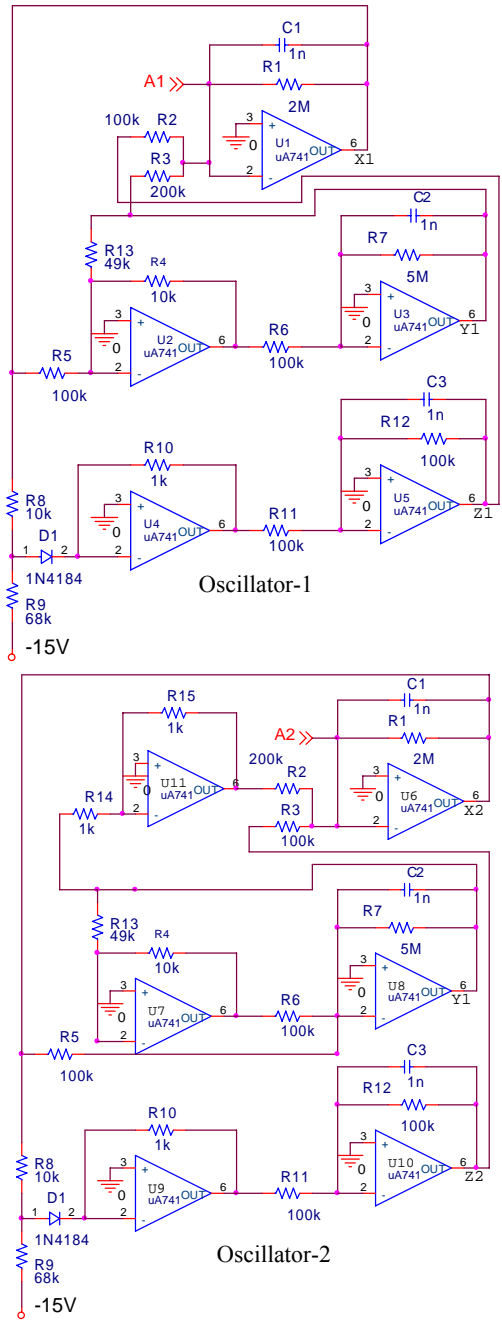

Fig.9. Counter-rotating piecewise linear Rossler circuits: upper one is oscillator-1 ($\omega=1$), lower one is oscillator-2 ($\omega=-1$). Components are denoted by same notations for easy reference in the text. However, their actual values are within 1% tolerances.

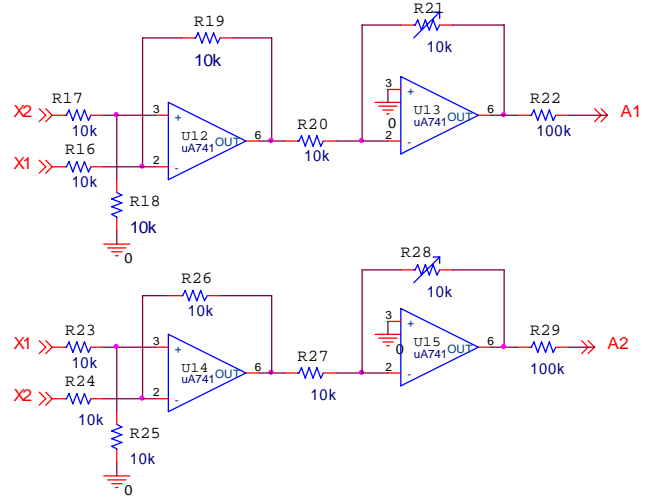

Fig.10. Coupling circuit for counter-rotating piecewise linear Rössler oscillators: $R_{21}$, $R_{28}$ determine the coupling strength.

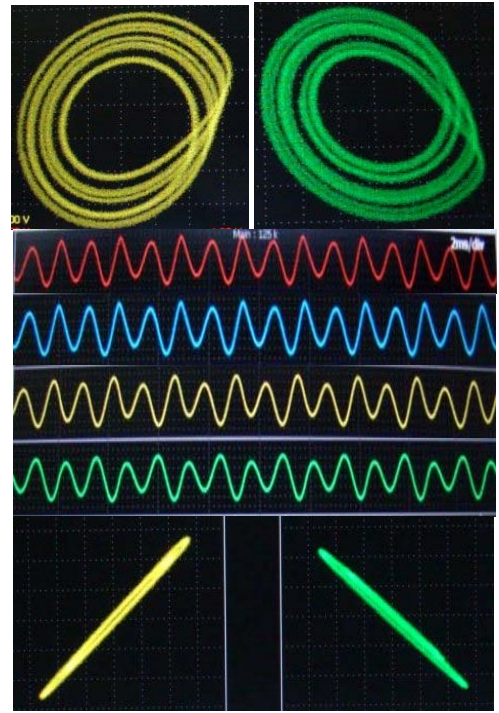

Fig.11. Oscilloscope pictures: counter-rotating attractors of piecewise linear Rössler model in upper row. Time series of $x_1$- and $x_2$-variable in second (red) and third (blue) rows respectively and they are in CS and. Time series of $y_1$- and $y_2$-variable in fourth (yellow) and fifth (green) rows respectively are in AS. *Lowest row*: $x_1$ vs. $x_2$ plot in CS (left) and $y_1$ vs. $y_2$ plot in AS (right).

Time series are actually measured as output voltages of three integrators as three state variables of each of the counter-rotating oscillators and then displayed in the oscilloscope. The experimental chaotic attractors (upper row) are clearly in opposite direction compared to each other. By connecting the coupling circuit, a MS scenario is obtained by properly tuning the resistance $R_{21}=R_{28}=R_F$, which decides the coupling strength. The MS state is realized for $R_F>5k\Omega$. The time series of $x_1$ and $x_2$ (online red and blue) shown in second and third rows respectively are in CS and, $y_1$ and $y_2$ (online yellow and green) in fourth and fifth rows respectively are in an AS state. The MS scenario is shown in the lower panel: CS ($x_1$ vs. $x_2$ plot) at left in yellow and AS ($y_1$ vs. $y_2$ plot) at right in green.

## VI. Conclusion

In conclusion, we reconfirmed the earlier report [4] that a MS scenario only emerged in counter-rotating oscillators under diffusive coupling. However, a general statement was still missing in ref.4 how to derive counter-rotating oscillators from a given dynamical model. We presented a general mathematical description how to induce counter-rotations in the trajectory of a dynamical system based on the Euler's rotation theorem [16]. Additionally, we clearly distinguished in a MS state that coexisting CS only emerged in the state variables those were related to the rotational plane and are directly coupled. The variables which were related to the rotational plane but not directly coupled, developed an AS state. We supported the results with numerical examples of limit cycle van der Pol system as well as chaotic models, Lorenz system and Rössler system. We established the stability criterion of the MS state in both the examples of chaotic models. Finally, we experimentally supported the results using electronic circuits of a limit cycle van der Pol oscillator and a piecewise linear Rössler oscillator in chaotic mode. We showed how to design counter-rotating electronic oscillators and then evidenced the onset of MS under scalar diffusive coupling. The MS is clearly found stable under intrinsic device noise and experimental noise for long run. However, the stability of MS in presence of external noise is not targeted here which is to be investigated in the future. The emerging collective behavior in a network of mixed population of counter-rotating oscillators is another important issue of our future research interest.

S.K.B. and S.K.D. acknowledge partial support by the BRNS, India, under grant # 2009/34/26/BRNS. S.K.D is grateful to Awadhesh Prasad, Manish Shrimali and Amit Sharma for very useful discussions.